\documentclass[a4paper,12pt]{article}
\usepackage[english]{babel}
\usepackage{amsmath,amssymb,amsfonts,amsthm}
\newtheorem{theorem}{Theorem}
\newtheorem{lemma}{Lemma}
\newtheorem{corollary}{Corollary}
\usepackage{graphicx}

\newcommand{\R}{\mathcal{R}}
\newcommand{\SB}{\partial \mathcal{B}}
\newcommand{\B}{\mathcal{B}}
\newcommand{\Rt}{\mathbb{R}^3}
\newcommand{\m}{\mathcal{M}}
\newcommand{\Si}{\Sigma}
\newcommand{\M}{M}
\newcommand{\J}{\mathcal{J}}

\newcommand{\E}{\mathcal{E}}

\title{The inequality between size and charge in spherical symmetry }
\author{Pablo Anglada, Sergio Dain, Omar E. Ortiz\\
  Facultad de Matem\'atica, Astronom\'{i}a y F\'{i}sica, \\
     Universidad Nacional de C\'ordoba, \\
Instituto de F\'{i}sica Enrique Gaviola, IFEG, CONICET,\\
  Ciudad Universitaria (5000) C\'ordoba, Argentina.}

\begin{document}
\maketitle

\begin{abstract}
  We prove that for a spherically symmetric charged body two times the radius
  is always strictly greater than the charge of the body. We also prove that
  this inequality is sharp. Finally, we discuss the physical implications of
  this geometrical inequality and present numerical examples that illustrate
  this theorem.
\end{abstract}

\section{Introduction}
Consider a body with angular momentum $J$ and electric charge $Q$. Let $\R$ be
a measure of the size of the body. The following inequality is expected to
hold for all bodies
\begin{equation}
  \label{eq:2v}
  \frac{Q^4}{4}+c^2 J^2 \leq  k^2 \frac{c^8}{G^2} \R^4,
\end{equation}
where $G$ is the gravitational constant, $c$ the speed of light and $k$ is an
universal dimensionless constant. These kinds of inequalities for bodies were
presented in \cite{Dain:2014xda}. They were motivated from similar kind of
inequalities valid for black holes (see the review article \cite{Dain:2014xda}
and references therein). The question of the ``minimum size'' for a charged
object (i.e. the case $J=0$) was first studied in \cite{arnowitt65}. Some
preliminary results were obtained in \cite{Anglada13b} for the case $Q=0$ and
in \cite{Rubio14} for the case $J=0$.

Heuristic physical arguments that support the inequality for the case $Q=0$
were presented in \cite{Dain:2013gma} and also, in that reference, a version of
this inequality was proved for constant density bodies, using a suitable
definition of size.  Khuri \cite{Khuri:2015zla} has proved it in a much more
general case, using the same measure of size as in
\cite{Dain:2013gma}. However, these inequalities are not expected to be sharp.

Recently Khuri \cite{Khuri:2015xpa} has proved a general version of inequality
in the case $J=0$ using a similar (but not identical) measure of size as the
one used in \cite{Dain:2013gma} and \cite{Khuri:2015zla}. As in the previous
case, this result is not expected to be sharp.

In these references the inequalities  have been studied in the two separated cases
$Q=0$ and $J=0$. The full inequality (\ref{eq:2v}) was presented for first time
in \cite{Dain:2015ira} using a completely different kind of heuristic arguments:
they are motivated by the Bekenstein bounds for the entropy of a body.  An
important property of the inequality (\ref{eq:2v}) is that there is only one
universal constant $k$ to be fixed.  Also, a rigidity statement for the
inequality (\ref{eq:2v}) was conjectured in \cite{Dain:2015ira}: the equality
is achieved if and only the entropy of the body is zero. In General Relativity,
this statement appears to imply that the equality can not be achieved for a
non-trivial body. 

The precise mathematical formulation of inequality (\ref{eq:2v}) involves
several difficulties. The most severe one is perhaps the definition of the size
$\R$ for a body in a general spacetime. An appropriate definition of $\R$ is
both difficult to find and non-unique.  Spherically symmetric spacetimes
represent an exception: the area radius of the boundary of the body is a
canonical definition for $\R$. The purpose of this work is to study inequality
(\ref{eq:2v}) in spherical symmetry (in particular, this implies $J=0$).  We
will prove several important properties of inequality (\ref{eq:2v}) which
currently can not be proved in a more general setting.  This will also allow us
to present the correct setting of the inequality in the general case.

First of all, we  determine the universal constant $k$ to be
\begin{equation}
  \label{eq:30}
  k=2.
\end{equation}
Secondly, we prove that inequality (\ref{eq:2v}) is sharp and strict: the
equality can not be achieved for a non-trivial body.  Moreover, the equality is
achieved in the asymptotic limit where the radius, charge and mass of the body
tend to zero. This is completely consistent with the argument presented in
\cite{Dain:2015ira}: the equality implies that the entropy of the body is
zero. In particular, a black hole can not reach the equality in (\ref{eq:2v})
since it has always a non-zero entropy and hence there is a gap between
inequalities for bodies and similar inequalities for black holes (which reach
equality for extreme black holes). This gap is given by a difference of a
factor $2$ in both inequalities. The existence of this gap is perhaps the most
relevant result presented in this article.

Finally, we prove that the correct setting for this inequality is an isolated
body that is not contained in a black hole. Inside a black hole, the inequality
can be violated. The appropriate definition for a body in this context is then:
a region of an asymptotically flat initial data that is not inside a horizon.

The plan of the article is the following. In section \ref{s:main} we present
our main result given by theorem \ref{teo} and we also discuss in detail it
physical implications.  In section \ref{s:proof} we prove theorem \ref{teo}. In
section \ref{s:numerical} we present numerical examples that illustrate the
assertions in theorem \ref{teo}. Finally, in appendix \ref{s:id} we summarize
useful properties of spherically symmetric initial data set. In the following we
use geometrized units where $G=c=1$.

\section{Main result}
\label{s:main}
The geometrical inequality between size and charge is appropriately formulated
in terms of an \emph{initial data set} for the Einstein equations.  For the
present results, we restrict ourselves to spherically symmetric initial data
where the 3-dimensional Riemannian manifold is taken to be $\Rt$. We call them
\emph{regular spherically symmetric} initial data. We also assume that the data
are \emph{asymptotically flat}. This kind of data has been extensively studied
in a series of articles by Guven and \'O Murchadha \cite{Guven:1994mt},
\cite{Guven:1994ms}, \cite{Guven:1997fc}.  In appendix \ref{s:id} we summarize
their basic properties and definitions.

Let  $\partial \B$ be a sphere centered at the origin with area radius
$\R$. That is, the area of $\partial \B$ is given by $4\pi \R^2$. The ball
enclosed by $\partial \B$ is denoted by $\B$. 

For a sphere $\partial \B$ we define the null expansions $\theta^+$ and
$\theta^-$ by \eqref{eq:expansions}.  A region between two concentric balls is
said to be \emph{untrapped} if $\theta^+\theta^->0$ on that region. The region
it is said to be \emph{trapped} if $\theta^+\theta^-<0$. The outer boundary of
a trapped region on an asymptotically flat data is called a \emph{horizon} and
it satisfies $\theta^+\theta^-=0$.  The area radius of the horizon is denoted
by $\R_0$.

\begin{theorem}
\label{teo}
Consider a regular spherically symmetric, asymptotically flat, initial data
set. Assume that there exists a ball $\B$ with finite radius $\R$ such that
outside $\B$ the data satisfy the electrovacuum constraint equations. Assume
also that in $\B$ the dominant energy condition holds.  Let $Q$ be the total
charge of $\B$, we assume $Q\neq 0$.  Then
 
\begin{itemize}
 \item[(i)] If  the exterior region outside $\B$ is untrapped, the inequality
 \begin{equation}
 \label{ineq1}
  2\R > |Q|,
 \end{equation}
 holds. 

\item[(ii)] If there is a horizon  outside $\B$, then the  radius
  $\R_0$ of the horizon satisfies the inequality
 \begin{equation}
 \label{ineq2}
  \R_0 \geq |Q|.
 \end{equation}
 The equality in (\ref{ineq2}) is achieved for the horizon of the extreme
 Reissner-Nordstr\"om black hole.
\end{itemize}

Moreover, we have:
\begin{itemize}
\item[(a)] The inequality (\ref{ineq1}) is sharp in the following sense: there
  exists a sequence of initial data that satisfy all the hypothesis of item (i)
  and such that in the limit the equality in (\ref{ineq1}) is achieved. In this
  limit, the radius, the charge and the total mass of this sequence tend to
  zero.

\item[(b)] The hypothesis of asymptotic flatness is necessary: there are
  examples of initial data which are not asymptotically flat but otherwise
  satisfy all the hypothesis in (i) for which the inequality (\ref{ineq1}) is
  violated.

\item[(c)] In the case (ii) there are examples where the radius $\R$ of the
  ball $\B$ (which is inside the horizon) violate the inequality (\ref{ineq1}).

\end{itemize} 

\end{theorem}
Let us discuss the scope and physical implications of this theorem. As it was
mentioned in the introduction, the original motivation to conjecture an
inequality of the form (\ref{ineq1}) for bodies comes from the analogous kind
of inequalities valid for black holes, namely, in our present setting,
inequality (\ref{ineq2}).  In reference \cite{Dain:2011kb} it has been shown
that inequality (\ref{ineq2}) is valid for general horizons (i.e. no symmetry
assumptions), it is a purely quasilocal inequality (i.e. no asymptotically flat
assumption is needed) and the equality is achieved for extreme black holes.
Since black holes are the ``most concentrated objects'' one would expect
naively that for fixed charge, the minimum possible radius in an inequality of
the form (\ref{ineq1}) is achieved for a black hole. Remarkably, theorem
\ref{teo} shows that it is not true: for fixed charge $Q$, the minimum possible
radius is $|Q|/2$ (and not $|Q|$ as in the case of a black hole). Example (a)
shows that this minimum radius is achieved in the asymptotic limit where the
radius, the charge and the total mass of the body (which is not inside a black
hole) tends to zero.  Non-trivial bodies always satisfy the strict inequality
(\ref{ineq1}). This is consistent with the discussion presented in
\cite{Dain:2015ira}: the equality in \eqref{ineq1} implies that the entropy of
the body is zero. Black holes (and also extreme black holes) have non-zero
entropy, hence there should be a gap between inequalities \eqref{ineq1} (for
bodies) and \eqref{ineq2} (for black holes), since the latter saturate for
extreme black holes. Theorem \ref{teo} shows that this gap is a factor $2$.

The canonical definition of radius in spherical symmetry is the areal radius
$\R$. There exists however another possible choice for the radius of a ball
$\B$: the geodesic distance to the center. But this radius has the disadvantage
that it can not be used, in general, for a black hole to obtain this kind of
inequalities. The black hole inequalities involve the area of the horizon or
quantities that depend, as the area, only on the geometry of the horizon (for
example, the shape of the horizon, see \cite{Reiris:2013jaa},
\cite{Reiris:2014tva}). The interior of the black hole does not appear to have
any physical meaning in this context. In particular, the geodesic distance and
also the radius used in \cite{Dain:2013gma} \cite{Khuri:2015xpa}
\cite{Khuri:2015zla} depend on the interior geometry of the body and hence, in
principle, they can not be applied to black holes. In theorem \ref{teo}, for
the fist time, the same radius definition is used for both bodies and black
holes. Finally we note that for some families of spherically symmetric initial
data it can be proved that the geodesic radius is greater than the areal radius
(see \cite{Bizon:1989xm} \cite{Guven:1994mt}) and hence for those cases,
inequality \eqref{ineq1} is also satisfied for the geodesic radius.

As we mention above, for a black hole the inequality \eqref{ineq2} can be
proved without using any asymptotic assumption. It depends only on the local
geometry near the horizon. This fact may suggest that a similar result can be
proved for a body $\B$. Namely, making hypothesis in the interior of the ball
$\B$ (regularity and dominant energy condition) and in a neighbourhood of the
boundary $\partial \B$ (the boundary is untrapped). However example (b) shows
that this is not possible.

Example (c) shows that inside a black hole the ball $\B$ with fixed charge $Q$
can be compressed to a radius $\R$ that violates the inequality
\eqref{ineq1}. And hence the hypothesis that the exterior region is untrapped
is necessary. Both examples (b) y (c) show that the correct setting for
inequality \eqref{ineq1} in general (i.e. without any symmetry assumption) is
the following: on an asymptotically flat initial data we consider a region that
is not contained in a black hole, this region is the appropriate definition of
``ordinary body'' in this context. These are precisely the hypotheses used in
the results presented \cite{Khuri:2015zla} and \cite{Khuri:2015xpa}.  We also
note that these hypotheses are required for the validity of the Bekenstein
bounds for the entropy (see \cite{Bekenstein:2004sh} \cite{Bousso:2002ju} and
reference therein).

In the spirit of the general results obtained in \cite{Khuri:2015zla} and
\cite{Khuri:2015xpa} about existence of black hole due to concentration of
angular momentum and charge, from theorem \ref{teo} we deduce the following
corollary.
\begin{corollary}
Consider a regular spherically symmetric, asymptotically flat, initial data
set. Assume that there exists a ball $\B$ with finite radius $\R$ such that
outside $\B$ the data satisfy the electrovacuum constraint equations. Assume
also that in $\B$ the dominant energy condition holds.  Let $Q$ be the total
charge of $\B$. If 
\begin{equation}
  2\R \leq |Q|,
 \end{equation}
then there are trapped surfaces enclosing $\B$.
\end{corollary}
Example (c) shows that this corollary is not empty. We will see that in this
example the data are not time symmetric and not maximal.

\section{Proof of theorem \ref{teo}}
\label{s:proof}
The proof is divided naturally in three parts, given by the following sections
\ref{s:part1}, \ref{s:part2} and \ref{s:part3}. The exterior region of the ball
is, by assumption, an asymptotically flat spherically symmetric solution of the
electrovacuum Einstein equations. Hence, by Birkhoff's theorem, this region is
described by the Reissner-Nordstr\"om metric which depends only on two
parameters: the mass and the charge.  This simple characterization of the
exterior region is the key simplification introduced by the assumption of
spherical symmetry. However, it turns out, that we do not need the full strength
of Birkhoff's theorem in the proof. We only need to compute the null expansions
of the spheres in term of the mass and the charge.  In section \ref{s:part1},
for the sake of completeness, we present a proof of this result. In the spirit
of theorem \ref{teo}, this proof is constructed purely in terms of the
constraint equations, in contrast with standard proof of Birkhoff's theorem
where the full Einstein equations are used.

In section \ref{s:part2} we prove the inequalities (\ref{ineq1}) and
(\ref{ineq2}). The key ingredient, introduced by Reiris in
\cite{Reiris:2014tva}, is the monotonicity of the Hawking energy (equivalent to
the Misner-Sharp energy in spherical symmetry) on untrapped regions.

Finally in section \ref{s:part3} we construct the three important examples (a),
(b) and (c). This examples are constructed using charged thin shells.

\subsection{The exterior region} 
\label{s:part1}
Consider  the constraint equations (\ref{eq:hamiltonean})--(\ref{eq:momentum})
in the exterior region of the ball $\B$. The electrovacuum assumption and the
spherical symmetry imply $j=0$, $\mu_M=0$ and $\rho=0$.  We first solve the Maxwell
constraint equations (\ref{maxwell}) in the exterior region, for the electric
field we obtain 
\begin{equation}
  \label{eq:7}
  E= \frac{Q}{r^2},
\end{equation}
where $Q$ is the total charge of the ball given by (\ref{eq:6}). For the
magnetic field we obtain a similar solution, but since we assume that there
are not magnetic charges the magnetic field vanishes. Then we have
\begin{equation}
  \label{eq:19}
  \mu = \frac{Q^2}{ 8 \pi r^4}, 
\end{equation}
and hence  the constraint equations (\ref{eq:hamiltonean})--(\ref{eq:momentum}) reduce to 
\begin{align}
 \label{c1}
  K_r \left( K_r +2 K_l \right) - \frac{1}{r^2} \left( {r'}^2 + 2 r r'' -1 \right)= \frac{Q^2}{r^4},     \\
 \label{c2}
   {K_r}' + \frac{r'}{r} \left( K_r - K_l \right) = 0.
\end{align}
From equation (\ref{c2}) we obtain
\begin{equation}
  \label{eq:8}
  r' K_l = \left( K_r r  \right)'.
\end{equation}
We multiply  equation  (\ref{c1}) by $r^4 r'$ and use the relation (\ref{eq:8})
to obtain
\begin{equation}
  \label{eq:9}
  r'r^4 {K_r}^2  +  r^4 2 K_r \left( K_r r \right)' - {r'}^3 r^2  - 2r^3 r' r'' + r' r^2 - r' Q^2 =0. 
\end{equation}
We rearrange terms in equation (\ref{eq:9}) to finally get
\begin{equation}
  \label{ce}
  \left( ({K_r} r)^2 - {r'}^2 \right) r^2 r' + \left(r^2 - Q^2\right) r'+
  \left( \left( \left( K_r r \right)^2 \right)'-2 r' r'' \right) r^3  =0. 
\end{equation}

Define the function $f(l)$ by
\begin{equation}
\label{deff}
 f=\frac{r^2}{4} \theta^+ \theta^- = {r'}^2 - \left( K_r r \right)^2,
\end{equation} 
where $ \theta^+$ and $ \theta^-$ are the null expansions  defined by (\ref{eq:expansions}). 
Note that the first term in (\ref{ce}) is proportional to $f$. We calculate $f'$
\begin{equation}
\label{f'}
 f'= 2 r' r'' - \left( \left( K_r r \right)^2 \right)'.
\end{equation}
We have that $f'$ it is proportional to the last term of (\ref{ce}). 
Then,  using (\ref{deff}) and (\ref{f'}) we  write (\ref{ce}) in the following form
\begin{equation}
  \label{eq:11}
   -r^2 r' f - r^3 f' + \left(r^2 - Q^2\right) r'=0
\end{equation}
We group the first two term in \eqref{eq:11} as a total derivative to finally obtain 
\begin{equation}
  \label{cea}
  -r^2  \left( f r \right)' +  \left(r^2 - Q^2\right) r' =0.
\end{equation}
Equation (\ref{cea}) can be integrated explicitly, the function $f$ is given by 
\begin{equation}
\label{f}
 f= 1 - \frac{2C}{r} + \frac{Q^2}{r^2},
\end{equation}
where $C$ is a constant.

Up to now, the calculations are local. If we assume that the exterior region is
asymptotically flat, then the constant $C$ that appears in the function $f$ is
the total mass (ADM mass) of the initial data. A simple way to obtain this
relation is by using the Misner-Sharp energy  defined by
\begin{equation}
  \label{eq:10}
  \E=\frac{r}{2}\left(1-\frac{r^2}{4}\theta^+ \theta^-\right).
\end{equation}
Using the definition of $f$ we write $\E$ in the form
\begin{equation}
\label{eq:Ef}
 \E  = \frac{r}{2} \left( 1- f \right) = C- \frac{Q^2}{2r}.
\end{equation}
From this expression we calculate the constant $C$ in terms of $\E$ and $Q$
\begin{equation}
 C= \E  + \frac{Q^2}{2r}.  
\end{equation}
A well known property of the energy $\E$ is that at infinity is equal to the
mass $M$ of the initial data (see \cite{Hayward:1994bu})
\begin{equation}
 M = \lim_{r \to \infty} \E.
\end{equation}
Then, taking this limit in equation (\ref{eq:Ef}) we finally obtain $C=M$, and
hence the final expression for $\E$ is give by
\begin{equation}
  \label{eq:ERN}
   \E  = M- \frac{Q^2}{2r}.
\end{equation}

We have computed the  product of the null expansions $\theta^+\theta^-$ 
in terms of the parameters $M$ and $Q$
\begin{equation}
  \label{eq:34}
 f=\frac{r^2}{4} \theta^+ \theta^-= 1 - \frac{2M}{r} + \frac{Q^2}{r^2}.
\end{equation}
This formula together with the formula for $\E$ given by (\ref{eq:ERN}) are the
only properties of the exterior region that will be used in the following steps
of the proof.

\subsection{The inequality}
\label{s:part2}
In this section we will prove the inequalities (i) and (ii). We have proved in
the previous section \ref{s:part1} that the product of the null expansions
(i.e. the function $f$ defined by (\ref{eq:34})) is characterized by only two
parameters: the mass $M$ and the charge $Q$.  We treat separately the cases
$M\geq |Q|$ and $M<|Q|$.

\subsubsection{$M\geq |Q|$ case}

Assume that the ball is located at the value $l_0$ of the geodesic distance to
center, that is $\R=r(l_0)$. The exterior region is defined by $r(l)$ with
$l\geq l_0$.

If $M \geq |Q|$, then $f$ has two real roots (or one double root
in the case of equality) at
\begin{equation}
  \label{eq:12}
 r_{+}=M + \sqrt{M^2 - Q^2}, \quad  r_{-}=M - \sqrt{M^2 - Q^2}.
\end{equation}
Note that $r_+\geq r_-$. 

For the exterior region we have two possibilities: either there exists at least
one point $l_1$ (with $l_1\geq l_0$) such that $r(l_1)=r_+$ or there is no such
a point. Consider the first case. Since $f=0$ at $r_+$, the exterior region is
not untrapped and hence we are in the case (ii) of the theorem. The 
horizon of the data is located as follows. If there is only one point $l_1$
such that $r_+=r(l)$, we take this point.  If there are many points that achieve
the value $r_+$ we take the most exterior one, i.e. if $r(l_1)=r(l_2)=r_+$ and
$l_1>l_2$, we take $l_1$. Let $l_1$ be such point. The asymptotic flatness
assumption implies that
\begin{equation}
  \label{eq:35}
  \lim_{l\to \infty} r(l)=\infty.
\end{equation}
Then $r(l)>r_+$ for $l>l_1$ (if not, this will contradict the assumption that
$l_1$ is the most exterior point with $r(l)=r_+$). And hence there are no
trapped surfaces in the region $ l>l_1$. Then, we have shown that $r(l_1)$ is
the horizon of the data.  The area radius of the horizon is $r_+$, hence we
have
\begin{equation}
  \label{eq:13v}
  \R_0=r_+=M + \sqrt{M^2 - Q^2}\geq |Q|.
\end{equation}
This proves the inequality (\ref{ineq2}) of theorem \ref{teo}. Note that for
extreme Reissner-Nordstr\"om (i.e. $M=|Q|$) the equality is achieved in
(\ref{eq:13v}).

Consider now the second case. If there are no points $l_1$, with $l_1\geq l_0$
such that $r(l_1)=r_+$, then by (\ref{eq:35}) we have that $r(l)> r_+$ for all
$l\geq l_0$. The exterior region is untrapped and we are in the case (i)
of theorem  \ref{teo}. We have proved that
\begin{equation}
  \label{eq:36}
  \R=r(l_0)> r_+ \geq |Q|.
\end{equation}
We emphasize that a stronger version of the inequality (\ref{ineq1}) is
satisfied for that case, since in (\ref{eq:36}) the factor $2$ is absent.

Note that in the previous argument we have not mentioned the radius $r_-$, but we
have used that $r_-\leq r_+$. For example, the ball $\B$ could be in the region
$0<r<r_-$ which is untrapped. However, since $r_-\leq r_+$ and we have
condition (\ref{eq:35}) in that case there will be always a point $l_1$ in the
exterior region such that $r(l_1)=r_+$.

\subsubsection{$M<|Q|$ case}
The case $M<|Q|$ is the most relevant one and it was proved by Reiris
\cite{Reiris:2014tva}. In what follows we essentially reproduce Reiris's proof.
The crucial ingredient is that the Misner-Sharp energy (\ref{eq:10}) is
monotonic on untrapped regions (see \cite{Hayward:1994bu}, \cite{Khuri:2009dt}). If we  assume that
on the region $l_1\leq l \leq l_2$ the dominant energy condition is satisfied
and $\theta^->0$, $\theta^+>0$, then
\begin{equation}
  \label{eq:24v}
  \E(l_1)\leq \E(l_2).
\end{equation}

We first prove the following result which is interesting by itself:
\begin{lemma}
  \label{l:ball}
  Consider a regular ball $\B$, such that the dominant energy condition is
  satisfied on $\B$.  If on the boundary $\partial \B$ of the ball $\B$ we have
  $\theta^->0$, $\theta^+>0$, then the Misner-Sharp energy of the boundary is
  non-negative
\begin{equation}
  \label{eq:37}
  \E(\SB)\geq 0. 
\end{equation}
\end{lemma}
Note that we are not assuming that the ball is embedded in an asymptotically
flat data, this is a quasilocal result that depends only on the interior of the
ball.

\begin{proof}
  Denote by $l_0$ the geodesic radius of the ball $\B$, that is $\R=r(l_0)$. To
  prove (\ref{eq:37}) we argue as follows. There are two cases: either the
  interior of $\B$ is untrapped or not. Consider the first case.  Since we have
  that $\theta^->0$, $\theta^+>0$ on the boundary, if the interior is untrapped
  (i.e. $\theta^+\theta^->0$) we obtain that $\theta^->0$, $\theta^+>0$ in
  $\B$.  It is well known that in the limit $l\to 0$ the Misner-Sharp energy is
  non-negative (see, for example, \cite{Szabados04} section 6.1.2). Since in
  the region $\B$ we have $\theta^->0$, $\theta^+>0$ we can use (\ref{eq:24v})
  with $l_1=0$ and $l_2=l_0$ to obtain
\begin{equation}
  \label{eq:56}
  0 \leq  \E(0)\leq \E(l_0).
\end{equation}
For the second case, we have, by assumption, that near the boundary
$\theta^+\theta^->0$. Hence, if the interior region of $\B$ is not untrapped
there should be a radius $r(l_1)$ in the interior of $\B$ such that
$\theta^+\theta^-=0$. From the expression (\ref{eq:10}) we have that the energy on
$r(l_1)$ is non-negative
\begin{equation}
  \label{eq:57}
  0\leq \E(l_1)=\frac{r(l_1)}{2}.
\end{equation}
In the region $l_1\leq l\leq l_0$ we have $\theta^->0$, $\theta^+>0$ and hence
we can use (\ref{eq:24v}) to obtain
\begin{equation}
  \label{eq:58}
  0\leq \E(l_1) \leq \E(l_0).
\end{equation}

\end{proof}

We continue with the proof. Note that since we have assumed $M<|Q|$ the
exterior region is untrapped, and hence we are in the case (i) of theorem
\ref{teo}. Moreover, since the data are asymptotically flat for large $r$ we
have that $\theta^+>0$ and $\theta^->0$ and hence, since the exterior region is
untrapped, we obtain $\theta^+>0$ and $\theta^->0$ in the whole exterior
region.  We can explicitly compute the Misner-Sharp energy $\E$ of the boundary
of the ball $\B$ using formula (\ref{eq:ERN}) and using lemma \ref{l:ball} we
obtain
\begin{equation}
 \E \left( \SB \right) = M - \frac{Q^2}{2\R} \geq 0.
\end{equation}
That is,
\begin{equation}
  \label{eq:13}
  \R  \geq \frac{Q^2}{2M} .
\end{equation}
We use that $M<|Q|$ to deduce from (\ref{eq:13}) the desired inequality 
\begin{equation}
  \label{eq:14v}
  2 \R \geq  Q.
\end{equation}

Finally, we prove that the inequality (\ref{eq:14v}) is strict, that is, no
material ball  can achieve the equality in (\ref{eq:14v}).  We argue by
contradiction.  Assume there exists a ball $\B$ such that $2\R = |Q|$. By
assumption, the exterior region is untrapped and hence the function $f$ is
positive on that region. We have two cases: $M \geq Q$ or $M<|Q|$. For the
first case we have already proved above that the stricter inequality
(\ref{eq:36}) is satisfied, and hence it is not possible to achieve $2\R = |Q|$
for that case. Consider the second case $M<|Q|$. We compute the energy $E$ at
the boundary
\begin{equation}
 \E \left( \SB \right) = M - \frac{Q^2}{2\R} = M-|Q| < 0,
\end{equation}
where we have used that $2\R = |Q|$. Then, the energy is negative and that 
contradicts lemma \ref{l:ball}.

\subsection{Examples}
\label{s:part3}
We construct in this section the examples (a), (b) and (c) of initial data
mentioned in theorem \ref{teo}. All the examples and much of the intuition
which led to the very formulation of theorem \ref{teo} were extracted from the
study of charged thin shells performed by Boulware \cite{boulware73}. In that
reference the complete dynamics of charged thin shells in the spacetime is
characterized. However, in this section we construct only initial data solving
the constraints in a self contained manner.  We make contact with the spacetime
picture just to favor the visualization.

We begin with the example (a). Consider the following spherically symmetric metric
\begin{equation}
  \label{eq:38}
  h = dl^2 + r^2(l)(d\theta ^2 + \sin^2 \theta d \phi ^2),
\end{equation}
where the radial function $r(l)$ is given by 
\begin{equation}
  \label{eq:25}
  r(l)=\begin{cases}
l  \quad \text{for } l\leq \R,\\
r_{RN}(l) \quad \text{for }  l\geq \R,
\end{cases}
\end{equation}
where $\R>0$ is an arbitrary constant and $r_{RN}(l)$ is the area radius
function corresponding to the Reissner-Nordstr\"om metric with mass $M$ and
charge $Q$. That is,  $r_{RN}(l)$ is the solution of the following differential equation
\begin{equation}
  \label{eq:39}
 r'_{RN}(l) =\left(1-\frac{2M}{r_{RN}} +\frac{Q^2}{r_{RN}^2} \right)^{1/2}.
\end{equation}
The integration constant in (\ref{eq:39}) is fixed by
the requirement $r_{RN}(\R)=\R$ and hence the function $r(l)$ defined by
(\ref{eq:25}) is continuous. 

The initial data set is prescribed with the metric (\ref{eq:38}) and zero
second fundamental form. The metric (\ref{eq:38}) describes a charged thin
shell of radius $\R$: the interior $l\leq \R$ is flat and the exterior is given
by the Reissner-Nordstr\"om metric. The metric depends on three parameters:
$(\R, M, Q)$.  But these parameters are not free if we imposes the dominant
energy condition on the metric. The dominant energy condition for time
symmetric data is equivalent to $R\geq0$, where $R$ is the scalar curvature of
the metric. To compute $R$ we first calculate the first and second derivatives
of the function $r(l)$ defined in (\ref{eq:25}). For the first derivative we
obtain
\begin{equation}
  \label{eq:41}
  r'(l)=\Theta(l-\R)\left( \left(1-\frac{2M}{r_{RN}} +\frac{Q^2}{r_{RN}^2} \right)^{1/2} -1 \right)+1,
\end{equation}
where $\Theta(x)$ is the step function defined by $\Theta(x)=0$ for $x<0$ and
$\Theta(x)=1$ for $x>0$. 
And for the second derivative we have
\begin{equation}
  \label{eq:42}
  r''(l)=\delta(l-\R) \left( \left(1-\frac{2M}{r_{RN}} +\frac{Q^2}{r_{RN}^2}
    \right)^{1/2} -1 \right) +\Theta(l-\R)\left( \frac{M}{r^2_{RN}}-\frac{Q^2}{r^3_{RN}}\right),
\end{equation}
where $\delta$ is the Dirac delta function. 

Using (\ref{eq:41}), (\ref{eq:42}) and the expression (\ref{eq:21}) for the scalar curvature
$R$ of the metric (\ref{eq:38}) we obtain 
\begin{equation}
  \label{eq:43}
  R=16\pi\sigma \delta(l-\R) +\Theta(l-\R) \frac{2Q^2}{r^4_{RN}}, 
\end{equation}
where we have defined
\begin{equation}
  \label{eq:44}
  \sigma= \frac{1}{4\pi l_0}\left(1- \left(
      1-\frac{2M}{\R}+\frac{Q^2}{\R^2}\right)^{1/2}  \right).
\end{equation}
The dominant energy condition $R\geq 0$ implies $\sigma \geq 0$, and this
impose restrictions on the value of the parameters. A convenient way to
express this relation is the following. Define the proper mass of
the shell by
\begin{equation}
  \label{eq:47}
  \m =4\pi \R^2 \sigma.
\end{equation}
Then, from (\ref{eq:44}) we obtain
\begin{equation}
  \label{eq:15}
  M=\m+\frac{Q^2-\m^2}{2\R}.
\end{equation}
The dominant energy condition is equivalent to $\m\geq 0$. 

To make contact with \cite{boulware73} we note that since the data are time
symmetric then the proper time derivative of the radius of the shell is zero in
the initial data and hence the 4-velocity of the shell ($u^\mu$ in the notation
\cite{boulware73}) is orthogonal to the spacelike hypersurface that define the
data. Then, using equations (\ref{eq:17x}) with $t^\mu=u^\mu$ we conclude that
$\sigma$ defined by (\ref{eq:44}) is identical to $\sigma$ defined by equation
(10) in \cite{boulware73}. And hence the proper mass $\m$ defined by
(\ref{eq:47}) is identical to the one defined in \cite{boulware73}. Note the
proper mass $\m$ is conserved along the evolution (see \cite{boulware73}). The
relation (\ref{eq:15}) is the special case of equation (16) in
\cite{boulware73} where the time derivative of the radius is zero.  We
emphasize that we have deduced the relation (\ref{eq:15}) using only the
dominant energy condition and the constraint equations. Expression
(\ref{eq:15}) was obtained for first time in \cite{arnowitt65}. In \cite{koc90}
this expression was generalized in the form of an inequality for spherical
distribution of charged matter momentarily at rest.

\begin{figure}
  \centering
   \includegraphics{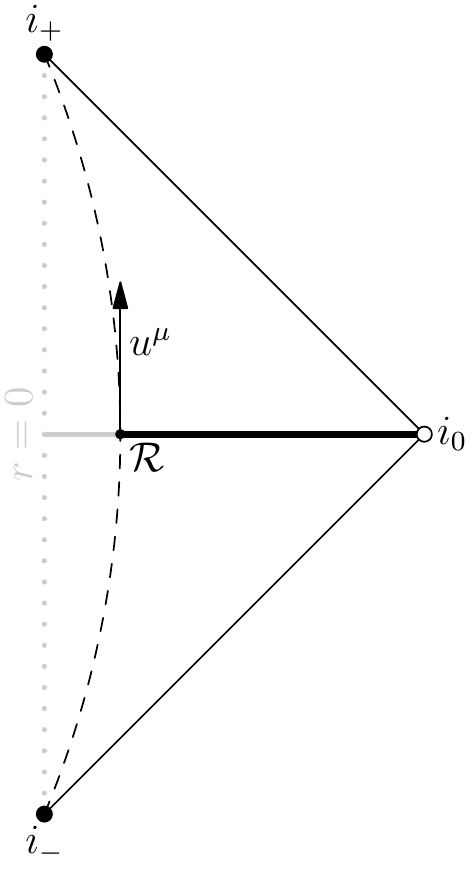} 
   \caption{The dashed line represents the trajectory of the shell. The shell
     has an infinite radius in the past $i_-$, it contracts to a minimum radius
     $\R$ and then it reexpands to infinite radius at $i_+$. The exterior
     region of the shell corresponds to the superextreme Reissner-Nordstr\"om
     spacetime. The interior region of the shell, drawn in gray, is flat. The
     spacelike surface of the initial data of Example (a) is represented by the
     thick horizontal line. The velocity of the shell is orthogonal to these
     initial data.}
  \label{fig:1}
\end{figure}

To construct the example (a) we will further impose that $M<|Q|$.  The
spacetime corresponding to these initial data is a shell that contracts to
a minimum radius $\R$ and then reexpands to infinity, see figure
\ref{fig:1}. The exterior region corresponds to the super-extreme
Reissner-Nordstr\"om spacetime.

The sequence of initial data is constructed as follows. We take the following
sequence of parameters, where $n\geq 1$ is a natural number
\begin{equation}
  \label{eq:16}
  \R_{n}=\frac{1}{n}, \quad Q_n=\frac{2}{n}-\frac{1}{n^2}, \quad
  \m_n=\frac{1}{2n^3}.  
\end{equation}
This sequence of initial data satisfies the dominant energy conditions since
$\m_n>0$. The total mass is computed using the formula (\ref{eq:15}), we obtain
\begin{equation}
  \label{eq:17}
  M_n=\frac{1}{n^3}+\frac{2}{n}-\frac{2}{n^2}-\frac{1}{8n^5}.
\end{equation}
Then we have
\begin{equation}
  \label{eq:18}
  M_n-Q_n=\frac{8n^2-8n^3-1}{8n^5}<0.
\end{equation}
There are no trapped surfaces in the exterior region and hence we are in
the case (i) of theorem \ref{teo}. Finally, we also have that
\begin{equation}
  \label{eq:26}
  \frac{Q_n}{2\R_{n}}=1-\frac{1}{2n}.
\end{equation}
From (\ref{eq:26}) we have that each member of the sequence satisfies the
inequality (\ref{ineq1}), as they should since the data satisfy the hypothesis
of the theorem for the case (i). Equation (\ref{eq:26}) implies that
the equality in (\ref{ineq1}) is achieved in the limit $n\to \infty$, and hence
we have proved that inequality (\ref{ineq1}) is sharp. Moreover, in the limit
$n\to \infty$ we have
\begin{equation}
  \label{eq:40}
  \lim_{n\to \infty} Q_n= \lim_{n\to \infty} \R_n= \lim_{n\to \infty} M_n=
  \lim_{n\to \infty} \m_n=0. 
\end{equation}

The second example (b) is constructed using the same metric (\ref{eq:38}), but
with different choice of parameters. We take $\m>0$ and
\begin{equation}
  \label{eq:27}
  |Q|>2\R.
\end{equation}
Using (\ref{eq:15}) and the assumption (\ref{eq:27}) we deduce that
\begin{equation}
  \label{eq:28}
 M>|Q|. 
\end{equation}
In addition, we take  $\R$ such that
\begin{equation}
  \label{eq:29}
  \R<r_-,
\end{equation}
where $r_-$ is given by (\ref{eq:12}).  Take $r_1$ such $\R< r_1 < r_-$ and we
consider the metric (\ref{eq:38}) defined up to $r_1$.

\begin{figure}
\centering
 \includegraphics[width=6cm]{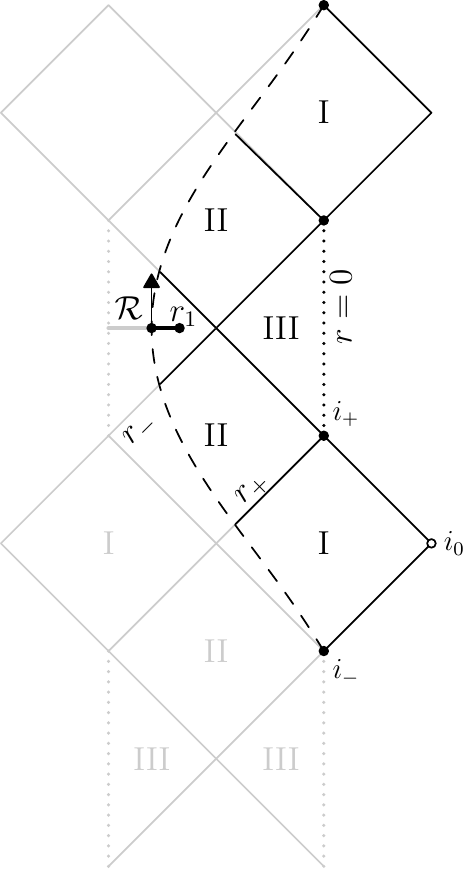} 
 \caption{The initial data of Example (b) is a piece of the time symmetric data
   located in region III of the Reissner-Nordstr\"om spacetime.}
\label{f:2}
\end{figure}

These data are, by construction, not asymptotically flat since they have a
boundary at $r_1$. The inequality (\ref{ineq1}) is not satisfied, since we have
imposed (\ref{eq:27}). In the exterior region of $\B$ up to $r_1$ there are no
trapped surfaces. These data are in region III of the Reissner-Nordstr\"om
spacetime, see figure \ref{f:2}. 

\begin{figure}
\centering
 \includegraphics[width=6cm]{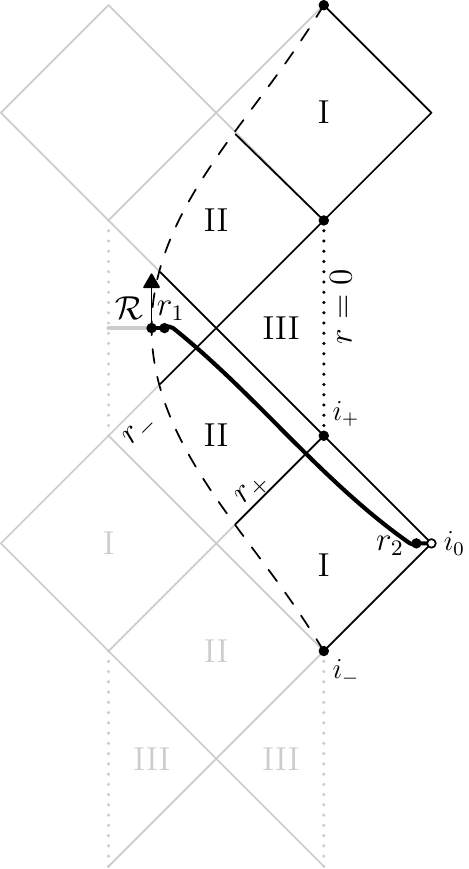} 
 \caption{Example (c) is constructed by extending the surface in Example (b) up
   to spacelike infinity $i_0$ in region $I$. The data are time-symmetric only
   in the regions $r<r_1$ and $r_2<r$.}
\label{f:3}
\end{figure}

Finally, we construct the third example (c). This example is based on the
previous example (b), but the data is extended to reach spacelike infinity. The
data are showed in figure \ref{f:3}. Note that these data are non-time
symmetric.
To construct the data we proceed as follows. Let $r_1$ and $r_2$ be two fixed
constants that satisfy $\R< r_1 < r_-<r_+<r_2$. The metric of the data is given
by (\ref{eq:38}) but now the function $r(l)$ is prescribed as follows
\begin{equation}
  \label{eq:25b}
  r(l)=\begin{cases}
l  \quad \text{for } l\leq \R,\\
r_{RN}(l) \quad \text{for }  l\geq \R,
\end{cases}
\end{equation}
where $r_{RN}(l)$ is a solution of the differential equation
\begin{equation}
  \label{eq:23}
  r'_{RN}(l)=\sqrt{f+(K_r r)^2},
\end{equation}
where $f$ is given by  (\ref{eq:34}) and the function $K_r$ is prescribed as follows. 
The function $f$  is negative in the region $r_-<r<r_+$.  Its minimum value 
\begin{equation}
  \label{eq:20}
  f_{min}=1-\frac{M^2}{Q^2},
\end{equation}
is achieved at the radius $r_{min}=Q^2/M$.  We prescribe the function $K_r(r)$
to be a smooth function with compact support in $[r_1,r_2]$ such that on the
interval $[r_-,r_+]$ it satisfies
\begin{equation}
  \label{eq:22}
  (K_r r)^2 > \frac{M^2}{Q^2}-1.
\end{equation}
Condition (\ref{eq:22}) ensures that the radicand on the right hand side of
(\ref{eq:23}) is always positive, hence
\begin{equation}
  \label{eq:45}
  r'_{RN}>0, 
\end{equation}
and we can integrate equation (\ref{eq:23}) to obtain a function $r_{RN}(l)$
which increases monotonously with $l$.  To complete the prescription of the data we
calculate the other piece $K_l$ of the second fundamental form using the
momentum constraint (\ref{c2}), that is
\begin{equation}
  \label{eq:24}
  K_l= \frac{r_{RN}}{r'_{RN}} K'_r+K_r.
\end{equation}
Note that equation (\ref{eq:24}) makes sense only if $r'_{RN}>0$.  We have constructed an
asymptotically flat initial data, such that there is an horizon in $r_+$ and
the inequality (\ref{ineq1}) is not satisfied by the ball $\B$. 
This finish the construction of example (c). 

Finally, it is interesting to mention the article \cite{Nakao:2013uj} where the
dynamics of two charged thin shells in spherical symmetry is analyzed. This
spacetime can provide more sophisticated examples that can have further
applications in the study of the inequality \eqref{ineq1}. For the particular
choice of parameters made in \cite{Nakao:2013uj} is simple to show that
inequality \eqref{ineq1} is satisfied. In the notation of \cite{Nakao:2013uj},
there are two concentric shells, the exterior one is called shell 2 and the
interior one shell 1. There are three regions: the exterior region $D_3$
outside shell 2, the region $D_2$ between shell 2 and shell 1 and the
interior region inside shell 1 $D_1$. It is assumed that in $D_3$ and $D_2$ the
spacetime is superextreme Reissner-Nordstr\"om (with parameters $(M_3,Q_3)$ and
$(M_2,Q_2)$ respectively) and in $D_1$ is Minkowski.
Clearly, Theorem \ref{teo} applies to shell 2 and not to shell 1. Also, since
in the exterior region $D_3$ the spacetime is superextreme
Reissner-Nordstr\"om, there are no trapped surfaces in $D_3$ and hence Theorem
\ref{teo} says that shell 2 should satisfy inequality \eqref{ineq1}.  However,
it turns out that due to the particular assumptions, the inequality
\eqref{ineq1} is also satisfied by shell 1. Let us explicitly prove these two
assertions.

The following condition should be satisfied at every shell  (see
\cite{Nakao:2013uj})
\begin{equation}
  \label{eq:1x}
  \E_{A+1}-\E_{A}>0,
\end{equation}
where $A=1,2$ and $\E_A$ denote the Misner-Sharp energy in the region $A$. 
Let us apply (\ref{eq:1x}) to shell 1. Since in $D_1$ the spacetime is
Minkowski we have $\E_1=0$  and hence we
obtain
\begin{equation}
  \label{eq:3x}
  \E_2>0.
\end{equation}
Using expression \eqref{eq:ERN} we obtain
\begin{equation}
  \label{eq:4x}
  \R_1 >\frac{Q_2^2}{2M_2},
\end{equation}
where $\R_1$ denotes the radius of shell 1. 
We use the assumption $M_2<|Q_2|$ on region $D_2$ to deduce from (\ref{eq:4}) the
desired inequality
\begin{equation}
  \label{eq:5xw}
  \R_1>\frac{|Q_2|}{2}.
\end{equation}
Now, we apply (\ref{eq:1x}) to the shell 2, we have
\begin{equation}
  \label{eq:6xw}
  M_3-\frac{Q^2_3}{2\R_2}>\E_2,
\end{equation}
and then
\begin{equation}
  \label{eq:7x}
  \R_2>\frac{\R_2\E_2}{2M_3}+\frac{Q^2_3}{2M_3},
\end{equation}
we use the assumption $M_3<|Q|_3$ on $D_3$ and equation (\ref{eq:3x}) to finally obtain
\begin{equation}
  \label{eq:8x}
   \R_2>\frac{|Q_3|}{2},
\end{equation}
where $\R_2$ denotes the radius of shell 2.

\section{Numerical examples}
\label{s:numerical}
In section \ref{s:part3} we have presented three important examples of initial
data that exhibit crucial properties of the inequality (\ref{ineq1}). These
examples are constructed in terms of charged thin shells and hence they have
distributional curvature. In this section we perform numerical computations of
initial data which have similar properties but they are generated by finite
smooth matter distribution. These computations are relevant for at least two
reasons. Firstly, for each example it will be clear that, changing slightly the
parameters, we obtain a whole family of data that shares the same
properties. That is, the examples are generic, they do not depend on a fine
tuning of the parameters. Secondly, the calculations presented here can have
further applications to test similar inequalities with different definition of
radius, like the one presented in \cite{Khuri:2015xpa}.

To solve the constraint equations (\ref{eq:hamiltonean})--(\ref{eq:momentum})
we proceed as follows. We use the momentum constraint (\ref{eq:momentum}) to
calculate $K_l$ as function of $K_r$ and $j$, namely
\begin{equation}
  \label{eq:48}
  K_l= \frac{r}{r'}K'_r+K_r- 4\pi \frac{r}{r'} j.
\end{equation}
Note that this equation makes sense only if $r'>0$. In all our examples with
$K_r\neq 0$ this
condition is satisfied.  Inserting (\ref{eq:48}) in the Hamiltonian constraint
(\ref{eq:hamiltonean}) we obtain
\begin{equation}
  \label{eq:49}
  3K^2_r+2 \frac{r}{r'}K'_rK_r-8\pi\frac{r}{r'} K_r j + \frac{1}{r^2}\left(r'^2+2rr''-1  \right)=8\pi \mu.
\end{equation}
In equation (\ref{eq:49}) we take the functions $K_r(l)$, $j(l)$ and $\mu(l)$
as free data and we solve for $r(l)$ imposing as initial conditions the
regularity conditions for the metric
\begin{equation}
  \label{eq:59}
  r(0), \quad r'(0)=1. 
\end{equation}

It is useful, for testing purposes, to have an integral expression for the
energy $\E$. This formula has been calculated in \cite{Guven:1994mt} and it is
given by
\begin{equation}
  \label{eq:54}
  \E= 4\pi \int_0^l dl \, r^2 \left( \mu r' + jrK_r \right).
\end{equation}

In our examples we impose 
\begin{equation}
  \label{eq:46}
  j=0,
\end{equation}
and we choose the  non-electromagnetic matter to vanish
\begin{equation}
  \label{eq:50}
  \mu_M=0.
\end{equation}
Then we have
\begin{equation}
  \label{eq:51}
  \mu=\frac{1}{8\pi} E^2.  
\end{equation}
The electric field must satisfy the Maxwell constrain equation (\ref{maxwell}).
We solve this equation as follows: we prescribe a smooth  function $Q(l)$ such
that at the origin $Q(l)=O(l^3)$ and it is constant for $l\geq l_0$ where $l_0$
represents the geodesic radius of the body.

Then our final equation is given by
\begin{equation}\label{ivp}\begin{split}
&r'' + \frac{1}{2r}\Bigl((r')^2 - 1\Bigr) = -\frac{Q^2}{2r^3} + \frac{3}{2} r
(K_r)^2 + \frac{r^2}{r'} K_r K'_r,\\
&r(0) = 0, \qquad \frac{ }{ } r'(0) = 1,
\end{split}
\end{equation}
where both $Q$ and $K_r$ are given functions of $l$. In \cite{Guven:1994ms} it
was observed that this initial value problem not only captures solutions
representing asymptotically flat initial data. If, for example, the charge is
concentrated enough around the origin then the solution $r(l)$ reaches a
maximum and returns to zero at finite geodesic distance. If, on the other
hand, $r$ grows big far away from the support regions of $K_r$ and the charge
density, then the forcing on the right hand side vanishes asymptotically and the
solution approaches $r'\simeq 1$ and $r''\simeq 0$, meaning asymptotic
flatness. Both of these behaviors will be shown in the numerical examples
below.

\subsection{The implementation} 

Equation \eqref{ivp} is a simple quasilinear ODE. It can
be written it as a first order system by defining $u=r$ and $v=r',$ 
\begin{equation}\label{ivps}
\left(\begin{array}{c}
u \\ v
\end{array}\right)^\prime = \left(\begin{array}{c}
v \\ \displaystyle{\frac{1-v^2}{2u} - \frac{1}{2u^3} Q^2(l) + \frac{3}{2} u K^2_r(l) +
\frac{u^2}{v}K_r(l)K'_r(l)}
\end{array}\right), 
\end{equation}
with initial condition
\begin{equation}\label{ivpsic}
\left(\begin{array}{c}
u(0) \\ v(0)
\end{array}\right) = \left(\begin{array}{c}
0 \\ 1
\end{array}\right).
\end{equation}
Now the geodesic distance $l$ can be discretized with a small step size $\delta
l$ and the problem solved with a standard ODE solver. We compute the numerical
solutions of \eqref{ivps}-\eqref{ivpsic} using the standard Runge-Kutta, 4th
order accurate, method.

We check the pointwise convergence of our code by computing a precision
quotient that depends on three numerical solutions to the same problem computed
using three different step sizes, $\delta l$, $2\delta l$ and $4\delta l$ (see
\cite{Kreiss-Ortiz-book}). This quotient should keep close, as a function of
$l$ and besides some isolated peaks, to the value $2^4$ if the code is correct
and the time step is small enough so that the truncation error is ${\cal
  O}(\delta l^4)$ for the three solutions.

A numerically computed solution will be a 4th order accurate approximation of
an exact solution if the latter is at least a $C^6$ smooth function of
$l$. This is so because the coefficient of the leading term in the truncation
error is proportional to the sixth derivative of the exact solution. To obtain a
solution $C^6$ smooth, one needs to prescribe a forcing which is $C^4$ smooth
as a function of $l$. To this end we introduce a monotonic polynomial, obtained
via Hermite interpolation
\begin{equation}
\label{polynomials}
 \begin{split}
p(a, b, x) &=  (1+w)^5(1 - 5 w +
15 w^2 - 35 w^3 + 70 w^4), \qquad w = \frac{x-b}{b-a},\\
q(c, d, x) &= 1 - p(c, d, x).
\end{split}
\end{equation}
For $a\le x\le b$, $p(a, b, x)$ is a monotonically increasing polynomial that
matches $0$ with $1$ in a $C^4$ smooth way. For $c\le x\le d,$ $q(c, d, x)$ is
a monotonically decreasing polynomial that matches $1$ with $0$ in a $C^4$
smooth way.

The energy integral (\ref{eq:54}) is approximated by a 4th order accurate
composite Simpson's rule. Also, as in the exterior region the energy and the
mass satisfy (\ref{eq:ERN}), we can compute the mass for any solution computed
on a finite $l$ interval that includes a portion of exterior region.

\subsection{Example (a)}

Here we compute the first few members of a sequence $\{r_n(l)\},~n=2, 3, \dots$
of regular solutions to the problem \eqref{ivp} that saturates the inequality
(3) in the limit $n\to \infty.$ This sequence must have the property that the
total charge $Q_n$ vanishes in the limit $n\to \infty,$ and consequently the
areal radius of the charge must also vanish in that limit, so that
$\lim_{n\to\infty} 2{\cal R}_n/Q_n = 1.$

All solutions in this sequence correspond to time symmetric initial data, that
is, in all this cases we set $K_r=0$ in the forcing of the equation
\eqref{ivp}.

We choose to compute the first few solutions of a sequence that satisfies
\begin{equation}\label{sequence}
Q_n = \frac{2}{n}, \quad \mbox{and}\quad {\cal R}_n = \frac{1}{n} +
\frac{1}{n\ln(n)}, \qquad n = 2, 3, 4, \dots
\end{equation}
This sequence of solutions is designed to saturate the inequality \eqref{ineq1}
in the limit $n\to \infty$ as
\begin{equation}\label{seq_convergence}
\frac{2{\cal R}_n}{Q_n} = 1 + \frac{1}{\ln(n)},
\end{equation}
with slow convergence to one.

Using the polynomial $p(a, b, x)$ defined in \eqref{polynomials} we prescribe
the function $Q(l)$ to be 
\begin{equation}
  \label{eq:52}
Q(l) =  \begin{cases}
Q_n p(a, l_0, l), & \mbox{if}~l< l_o,\\
Q_n , & \mbox{if}~l \ge l_0,
\end{cases}\qquad a=0,\quad l_0>0,\quad Q_n = \frac{2}{n},  
\end{equation}
where $l_0$ is the geodesic radius of the charge distribution.  At the origin
the function $Q(l)$ vanishes as ${\cal O}(l^5)$.

To compute each solution of the sequence, say with index $n$, the value
of the total charge $Q_n$ and the geodesic radius $l_0$ of the charge are input
parameters in the program. The areal radius of the charge $\R(l_0)$
is known only after the solution is computed. Thus, the input parameter $l_0$
needs to be adjusted to obtain the desired value $\R(l_0) = \R_n.$ 
To adjust $l_0$ we start with two solutions with the right charge, one with
smaller value of $\R$ and another with larger value of $\R$. We then
perform a bisection procedure on  $l_0$ to find the root of the
function
\begin{equation}
  \label{eq:53}
g(l_0) = \R(l_0) - \frac{1}{n} - \frac{1}{n\ln(n)}. 
\end{equation}
We stop the iterations when the value of $\R(l_0)$ reaches the value of
$\R_n$ with ten correct digits. Table \ref{table_sequence} shows the relevant
input parameters we obtain for the first few members of the sequence of
solutions and the  mass that results for each of them.

\begin{table}[t]
\begin{center}
{\small
\noindent
\begin{tabular}{ccrlcc}
\hline\hline
$n$ & $Q_n$ & $\delta l$\hspace{1.7em} & \hspace{5em}$l_0$ & mass \\
\hline\hline
2 & 1   & $1\times 10^{-3}$ & 1.346158647537232                 & 0.680983 \\
3 & 2/3 & $1\times 10^{-3}$ & 7.422593683004379$\times 10^{-1}$ & 0.554538 \\
4 & 1/2 & $5\times 10^{-4}$ & 5.176483931019902$\times 10^{-1}$ & 0.449646 \\
5 & 2/5 & $5\times 10^{-4}$ & 3.981155012268573$\times 10^{-1}$ & 0.375407 \\
6 & 1/3 & $5\times 10^{-4}$ & 3.235192440450192$\times 10^{-1}$ & 0.321540 \\
7 & 2/7 & $2\times 10^{-4}$ & 2.724420906044543$\times 10^{-1}$ & 0.280995 \\
8 & 1/4 & $1\times 10^{-4}$ & 2.352533040568233$\times 10^{-1}$ & 0.249473 \\	
\hline\hline
\end{tabular}
\caption{Parameters and mass for the first solutions in the sequence
satisfying \eqref{sequence}.}\label{table_sequence}
}
\end{center}
\end{table}

\begin{figure}[h]
\begin{center}
\includegraphics[width=0.6\columnwidth]{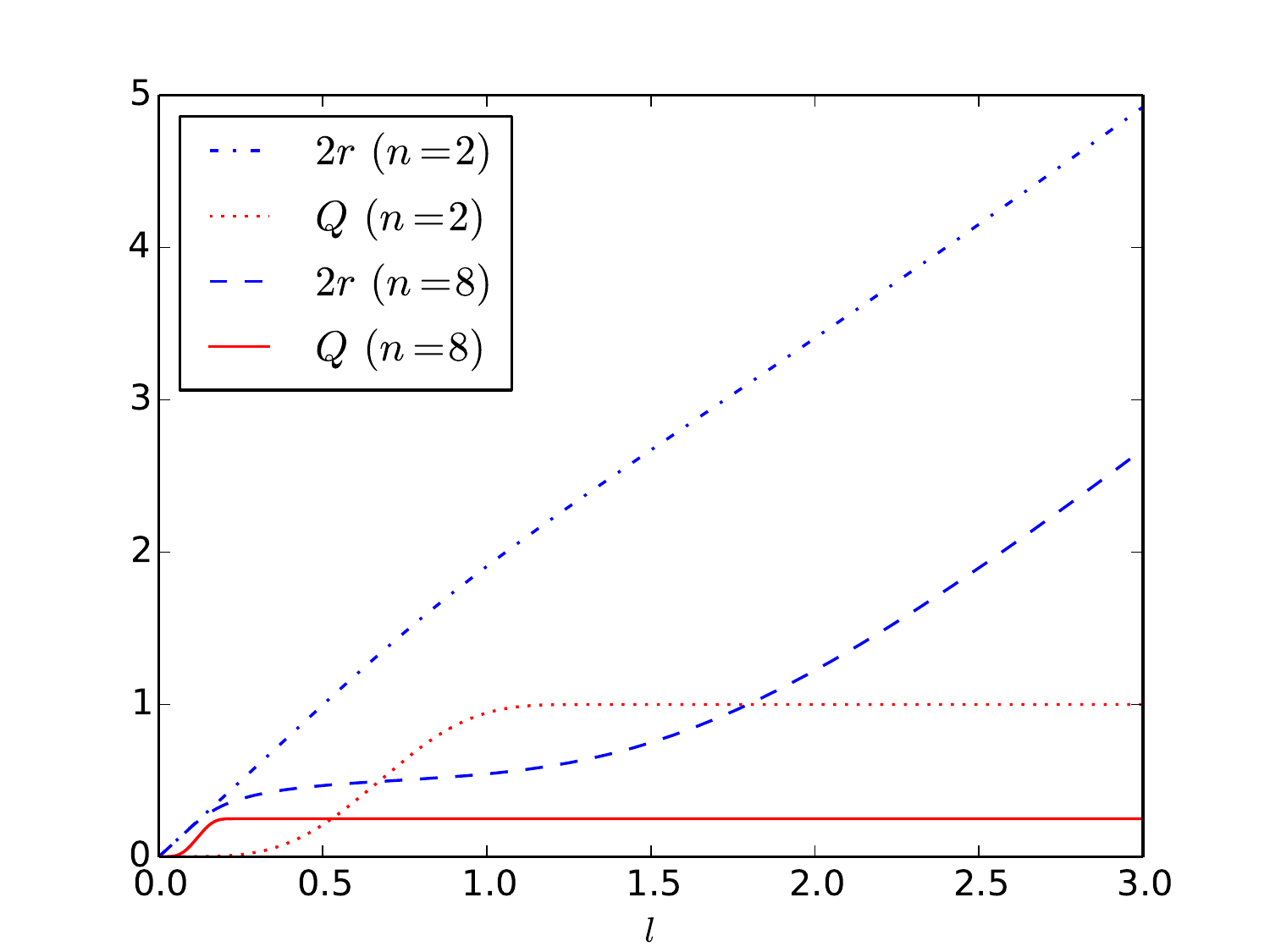}
\caption{$2r(l)$ and $Q(l)$ for the solutions with $n=2$ and $n=8$. The border
of the objects are placed at the corresponding values of $l_0$ given in Table
\ref{table_sequence}.}\label{fig_seq_1}
\end{center}
\end{figure}

\begin{figure}[h]
\begin{center}
\includegraphics[width=0.6\columnwidth]{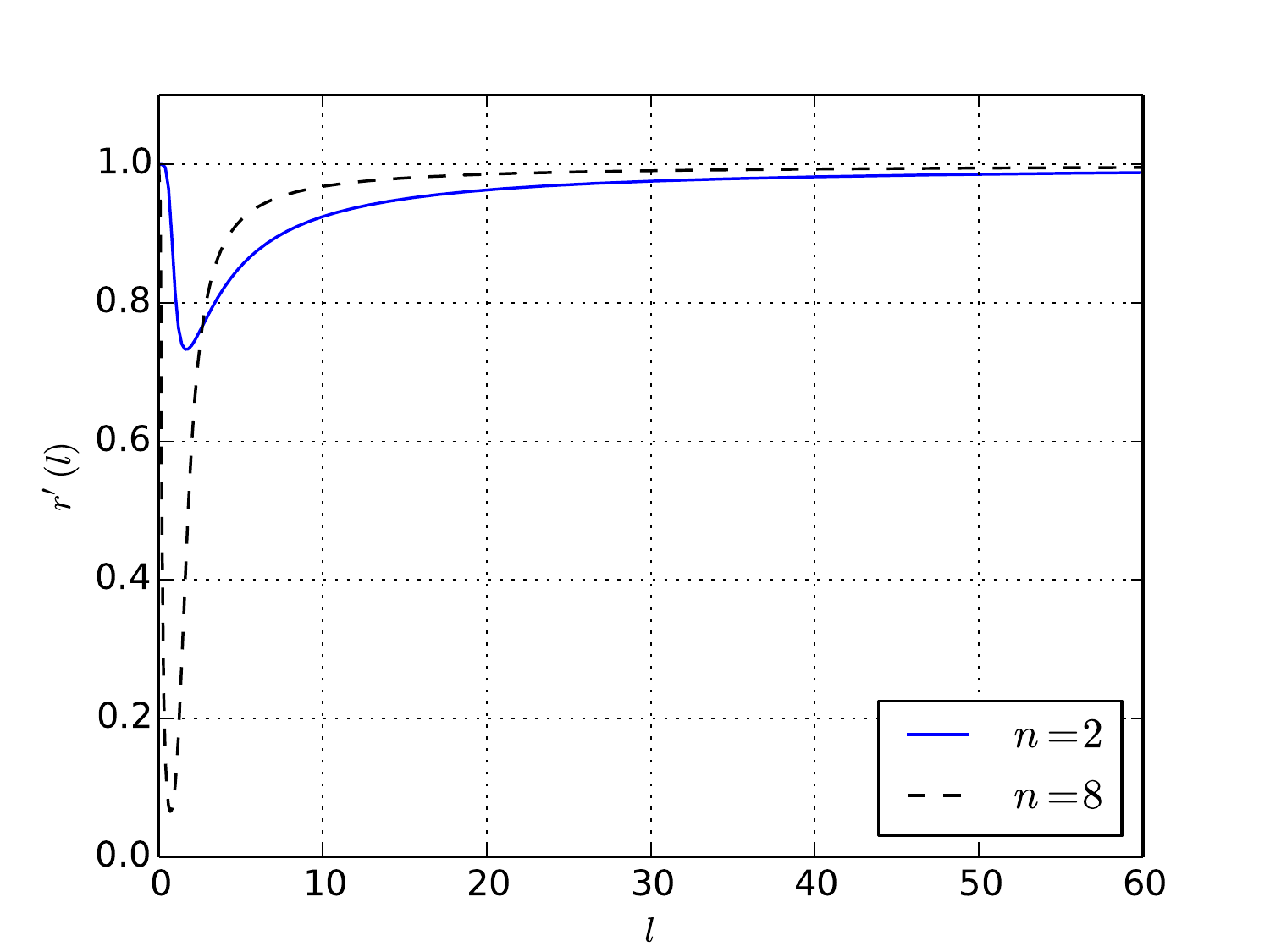}
\caption{Plots of $r'(l)$ for the solutions with $n=2$ and $n=8$ of Table 1 (showing asymptotic flatness).}\label{fig_seq_2}
\end{center}
\end{figure}

To illustrate the behavior of the solutions in this sequence two plots are
shown. Figure \ref{fig_seq_1} shows the plots of $2r(l)$ and $Q(l)$ of the
first ($n=2$) and last ($n=8$) solutions in Table \ref{table_sequence} in a
small region around the charge domain. Figure \ref{fig_seq_2} shows the plots
of $r'(l)$ for all solutions in Table \ref{table_sequence} in a larger region.
These last plots show how the solutions satisfy the asymptotic boundary
condition. Note that $|r'|\leq 1$, this is always true for time symmetric initial data, see \cite{Guven:1994ms}. 

\subsection{Example (b)}

\begin{figure}[h]
\begin{center}
\includegraphics[width=0.6\columnwidth]{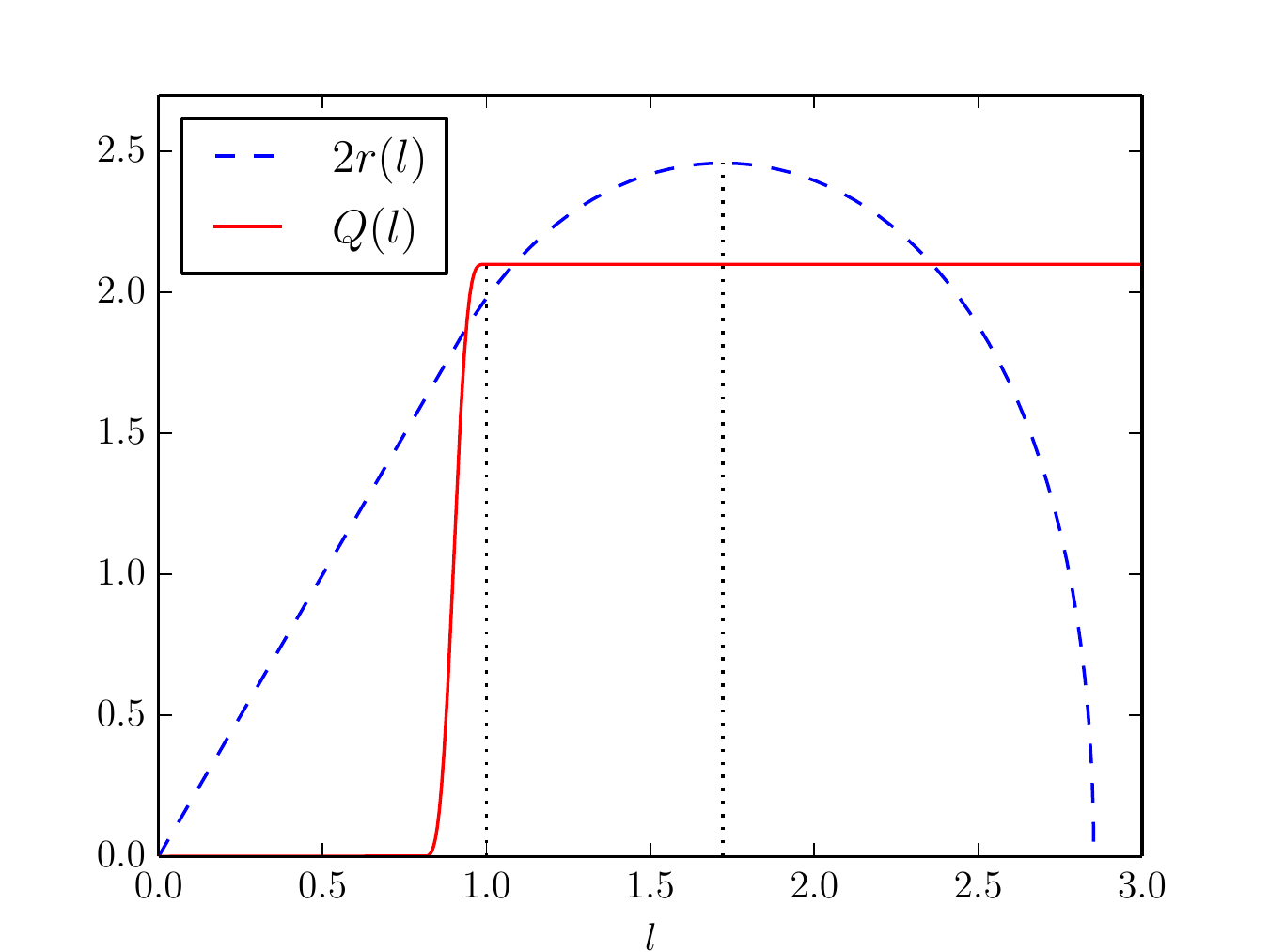}
\caption{ $2r(l)$ and $Q(l)$ for the solution of example (b), which is not asymptotically flat. The vertical dotted lines indicate the values of $l_0 = 1.0$ (the border of the body) and $l_1 = 1.22959$ where $r'$ becomes zero. The inequality (3) is violated by about $6\%$.}
\label{fig_example_b}
\end{center}
\end{figure}

In this section we present a single numerical solution representing time
symmetric initial data. The charge distribution is a thick spherical shell with
support in a finite interval $0 < a \le l \le l_0$. The charge $Q(l)$ is given by
\begin{equation}\label{Q_example_b}
Q(l) =  \begin{cases}
0, & \mbox{if}\quad l\le a,\\
Q p(a,b,l), & \mbox{if}\quad a< l < l_0,\\
Q, & \mbox{if}\quad l_0 \le l,
\end{cases}\qquad a=0.8,~l_0=1.0,~Q = 2.1 .
\end{equation}
The solution with these parameters violates the inequality \eqref{ineq1}; the
total charge $Q$ exceeds $2{\cal R}$ by more than 6\%. Figure
\ref{fig_example_b} shows a plot of this solution. At about $l=2.85200$, $r(l)$
gets back to zero.  At this point the equation becomes singular and the
solution diverges. As expected $r'(l)$ vanishes outside the body (maximum of
$r(l)$) at about $l_1=1.72169$, with $r(l_1) = 1.229588$, showing that there
exist a trapped surface enclosing the body. However, near the boundary of the
body (i.e. in the region $l_0\leq l < l_1$) there are no trapped surfaces.

As a test for the solution, using formula (\ref{eq:ERN}) we calculate the mass
$M= 2.408077371$ and then we calculate $r_-$ given by (\ref{eq:12}). The value
of $r_-$ coincides with the value $r(l_1)$ calculated above with seven digits.

\subsection{Example (c)}

\begin{figure}
  \centering
  \includegraphics[width=0.6\columnwidth]{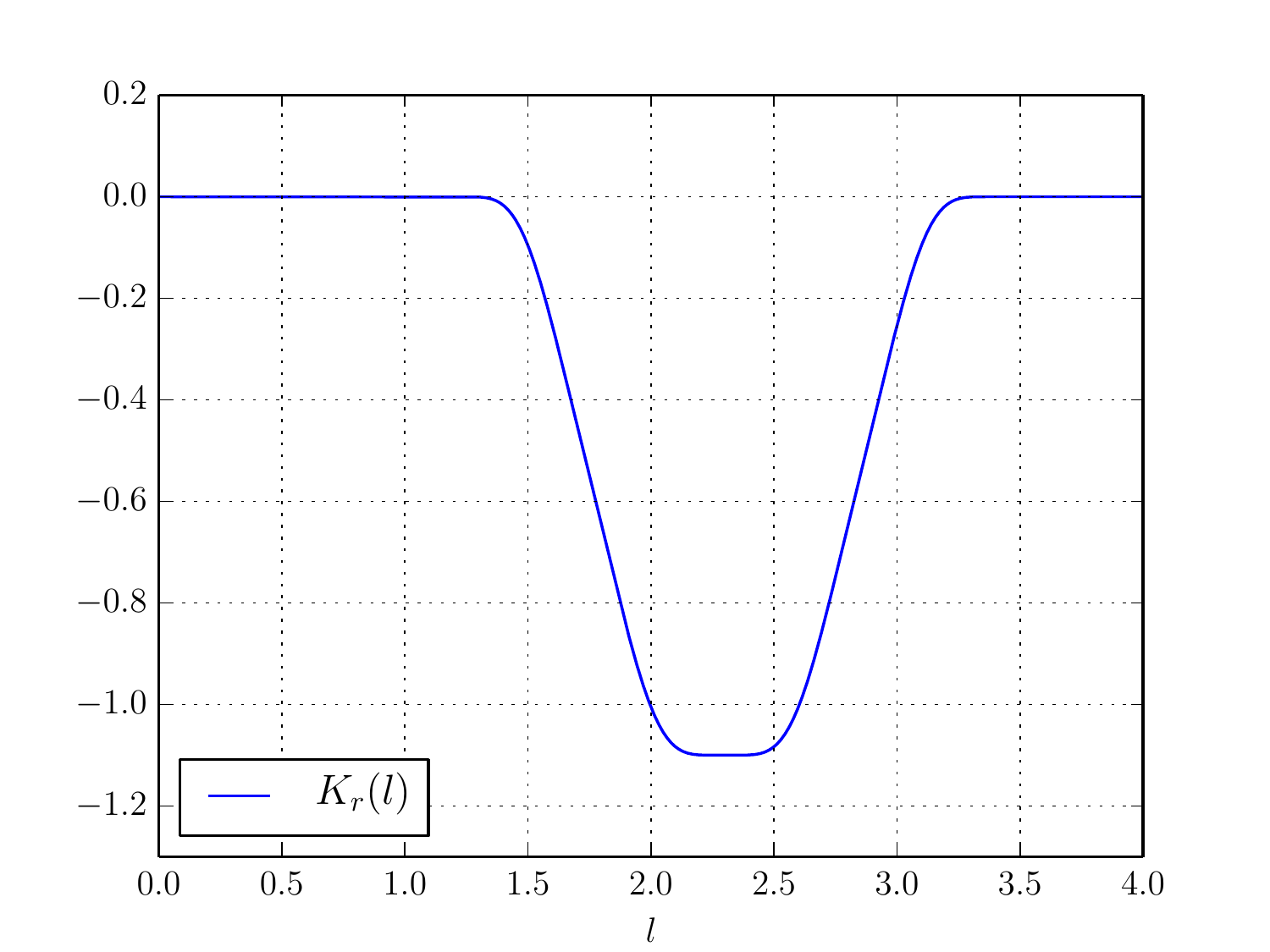}
  \caption{Plot of $K_r(l)$, $C^5$ smooth with compact support in $[1.2, 3.4]$.}
  \label{fig:Kr}
\end{figure}

In this section we modify the data used to obtain the solution of Example (b).
This is done as suggested by the analytical examples of section \ref{s:part3}.
The charge distribution is the same as in example (b), so that $Q(l)$ is given
by \eqref{Q_example_b}, but now there is a non-vanishing extrinsic curvature
$K_r(l)$ of compact support, thus the solution no longer represents time
symmetric initial data. We prescribe $K_r'(l)$ as the $C^4$ smooth function
\begin{equation}
\label{K_example_c}
\begin{split}
K_r'(l) = (-2.0)\times\begin{cases}
0, & \mbox{if} \quad 1.2 \le l,\\
p(1.2, 1.75, l), & \mbox{if} \quad 1.2 < l \le 1.75,\\
q(1.75, 2.3, l), & \mbox{if} \quad 1.75 < l \le 2.3,\\
-p(2.3, 2.85, l), & \mbox{if} \quad 2.3 < l \le 2.85,\\
-q(2.85, 3.4, l), & \mbox{if} \quad 2.85 < l < 3.4,\\
0, & \mbox{if} \quad 3.4 \le l,\\
\end{cases}
\end{split}
\end{equation}
where $p$ and $q$ are the polynomials defined in \eqref{polynomials}. $K_r(l)$
is defined as the exact integral of $K_r(l)$. In figure \ref{fig:Kr} we show a plot of $K_r$.

\begin{figure}[h]
\begin{center}
\includegraphics[width=0.6\columnwidth]{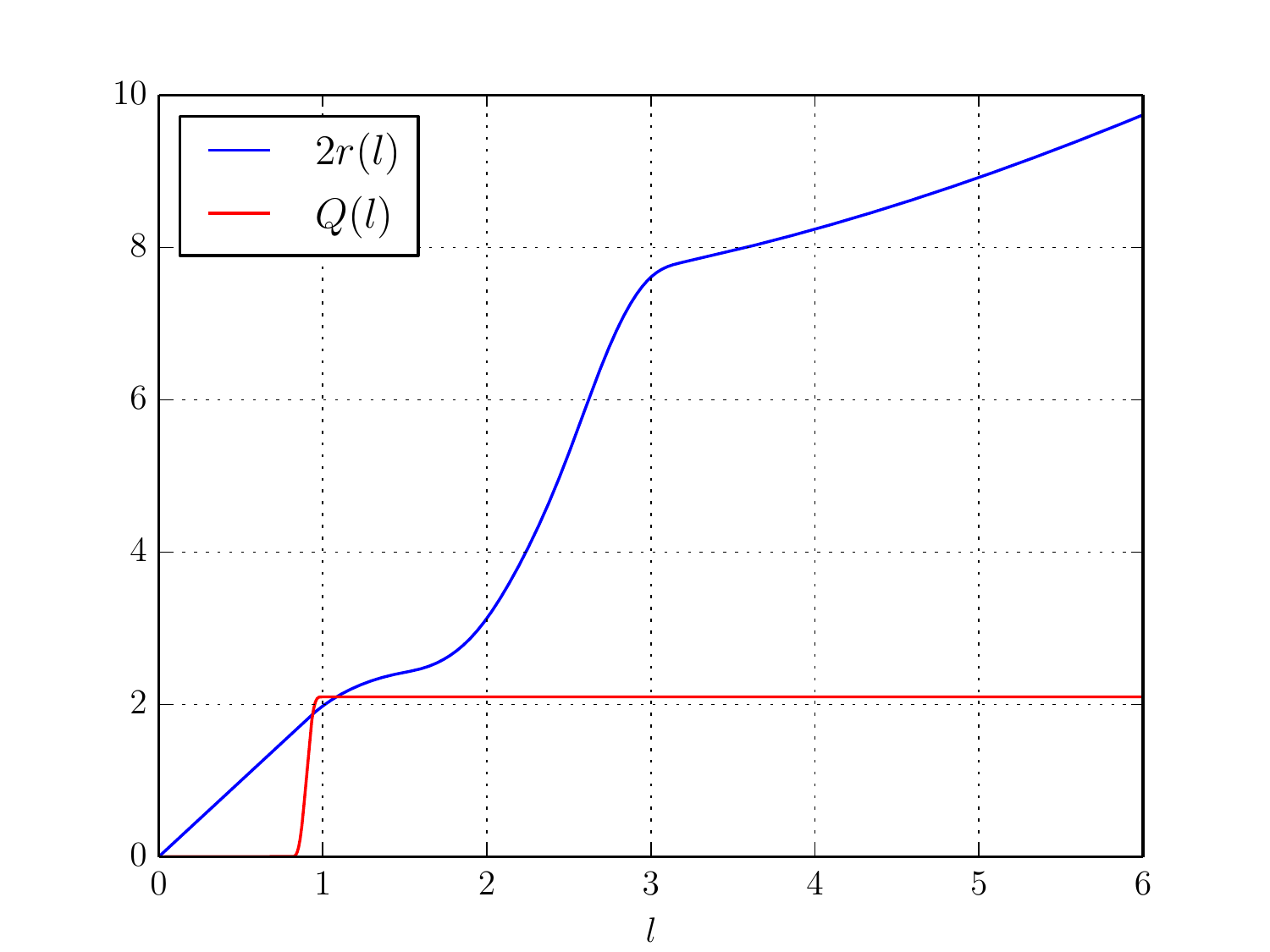}\\
\includegraphics[width=0.6\columnwidth]{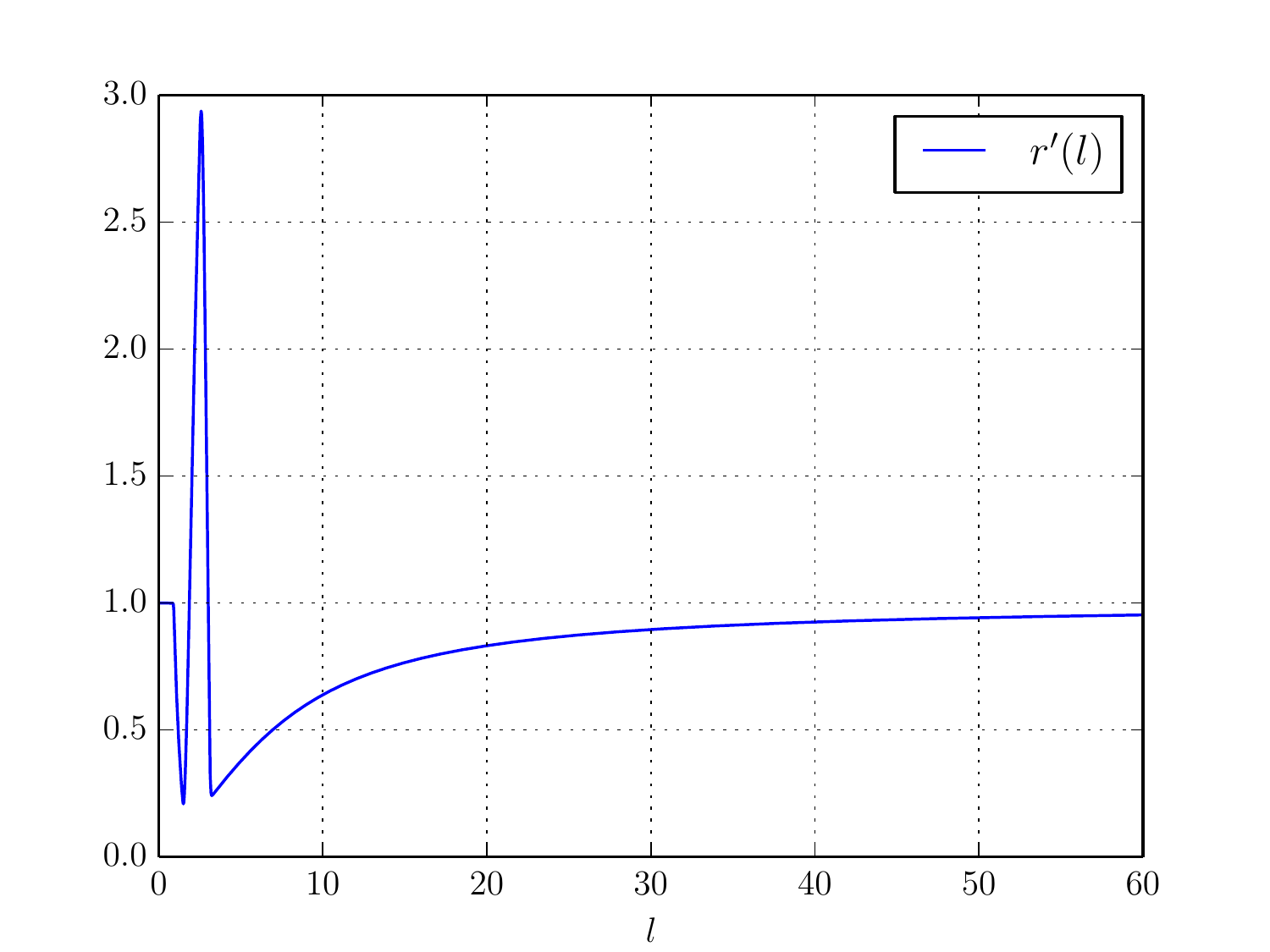}
\caption{$2r(l)$ and $Q(l)$ (upper plot), $r'(l)$ on a larger domain (lower
plot), for the solution obtained with \eqref{Q_example_b} and
\eqref{K_example_c}.}\label{fig_example_c_1}
\end{center}
\end{figure}

The solution obtained is a monotonically increasing $r(l)$ coincident with
the solution of example (b) when $l\le 1.2$ (the initial value problem is
exactly the same up to this point). For larger values of $l$ the extrinsic
curvature affects the solution so that $r(l)$ keeps growing and the solution
becomes asymptotically flat. Figure \ref{fig_example_c_1} shows the behavior of
this solution.

\begin{figure}[h]
\begin{center}
\includegraphics[width=0.6\columnwidth]{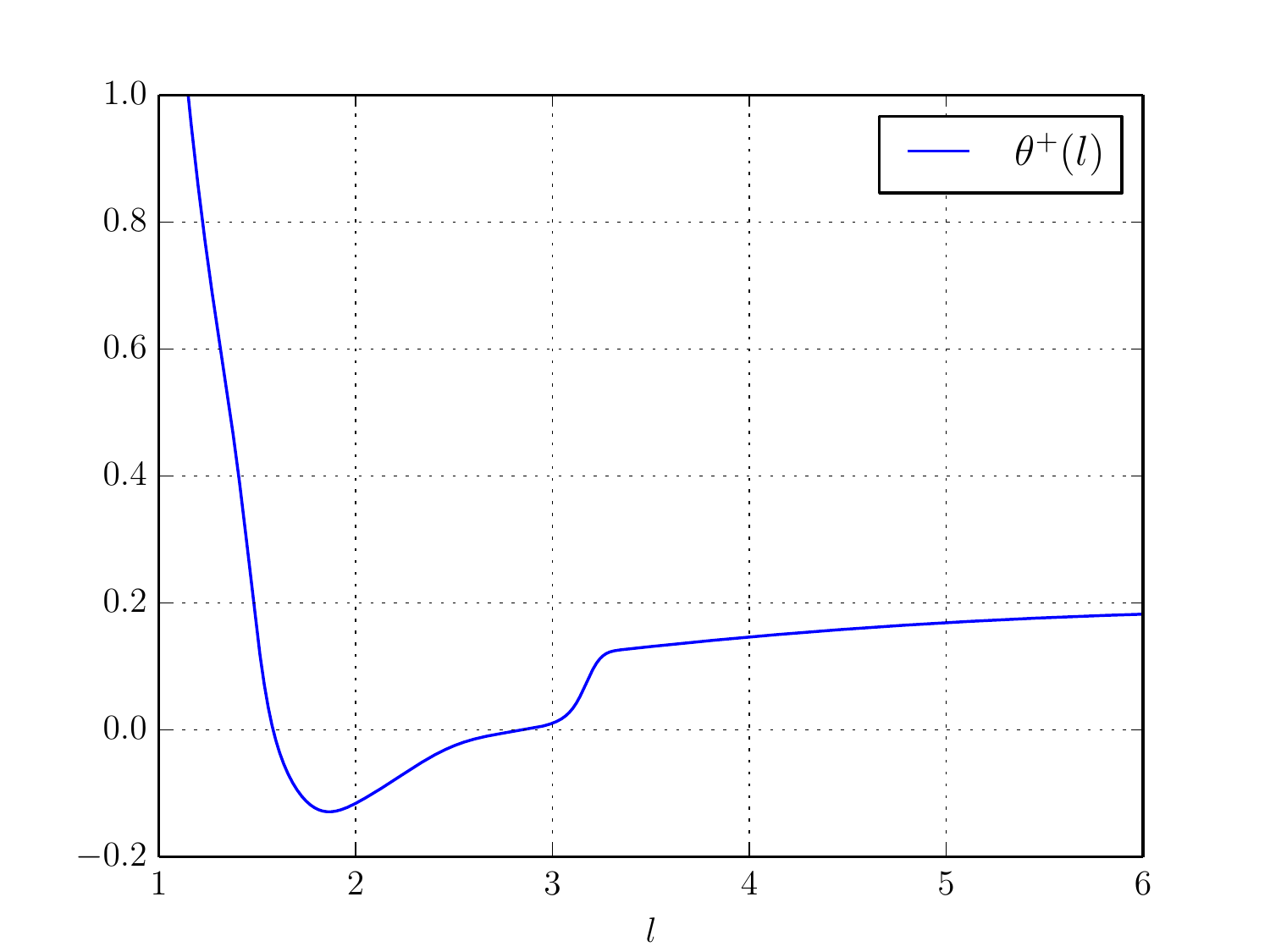}
\caption{Plot of $\theta^+(l)$ for the solution of example
(c).}\label{fig_example_c_4}
\end{center}
\end{figure}

This solution has a horizon outside the body. Figure \ref{fig_example_c_4} shows
the plot of $\theta^+(l).$ This function has two roots located at
$l^-~=~1.58085$ and $l^+~=~2.85231$. These values correspond to radii $r(l^-) =
1.22959$ and $r(l^+) = 3.58657$ respectively. The computed  mass for this
solution is $M=2.408077371$. The total charge, $Q=2.1$, is an input parameter in
the program. We can compute the values $r_-$ and $r_+$ given by equation \eqref{eq:12},
which turn out to be coincident with the values $r(l^-)$ and $r(l^+)$ in seven
and six digits respectively. The radius of the horizon, ${\cal R}_0=
r_+$ clearly satisfies the inequality \eqref{ineq2}.
Finally, using formula (\ref{eq:48}) we numerically compute $K_l$ and then we
compute the trace of the second fundamental form given by $K=K_l+2K_r$. This
function is non zero, and hence the data are not maximal.

\appendix

\section{Spherically symmetric initial data for the Einstein-Maxwell equations}
\label{s:id}

Let $\M$ be a  4-dimensional manifold with metric $g_{\mu\nu}$ (with
signature $(-+++)$) and Levi-Civita connection $\nabla_\mu$. In the following,
Greek indices $\mu, \nu \cdots$ are always 4-dimensional.

Consider Einstein equations with energy momentum tensor $T_{\mu\nu}$
\begin{equation}
  \label{eq:3a}
  G_{\mu\nu}=8\pi T_{\mu\nu},
\end{equation}
where $G_{\mu\nu}$ is the Einstein tensor of the metric $g_{\mu\nu}$.
 The \emph{dominant energy condition} for $T_{\mu\nu}$  is given by 
\begin{equation}
  \label{eq:15v}
  T_{\mu\nu} v^\mu w^\nu \geq 0,
\end{equation}
for all future-directed causal vectors $v^\mu$ and $w^\nu$. 

It will be useful to decompose the matter fields   $T_{\mu\nu}$ into the electromagnetic part 
and the non-electromagnetic part
\begin{equation}
  \label{eq:14}
  T_{\mu\nu}=T^{EM}_{\mu\nu}+T^M_{\mu\nu},
\end{equation}
where $T^{EM}_{\mu\nu}$ is the electromagnetic energy momentum tensor given by
\begin{equation}
  \label{eq:4}
  T^{EM}_{\mu \nu}= \frac{1}{4\pi}\left(F_{\mu \lambda}
    F_\nu{}^{\lambda}-\frac{1}{4}g_{\mu \nu} F_{\lambda \gamma} F^{\lambda \gamma}  \right),
\end{equation}
and $F_{\mu\nu}$ is the (antisymmetric) electromagnetic field tensor
that satisfies Maxwell equations
\begin{align}
  \label{eq:maxwell}
  \nabla^\mu F_{\mu\nu} & =-4\pi \J_\nu, \\ 
\nabla_{[\mu} F_{\nu \alpha]} & =0,
\end{align}
where $ \J_\nu$ is the electromagnetic current.

Initial conditions for Einstein equations are characterized by \emph{initial
  data set} given by $(\Si, h_{ij}, K_{ij}, \mu, j^i)$ where $\Si$ is a
connected 3-dimensional manifold, $h_{ij} $ a (positive definite) Riemannian
metric, $K_{ij}$ a symmetric tensor field, $\mu$ a scalar field and $j^i$ a
vector field on $\Si$, such that the constraint equations
\begin{align}
 \label{const1}
   D_j   K^{ij} -  D^i   K= -8\pi j^i,\\
 \label{const2}
   R -  K_{ij}   K^{ij}+  K^2=16\pi \mu,
\end{align}
are satisfied on $\Si$. Here $D$ and $R$ are the Levi-Civita
connection and scalar curvature associated with $ {h}_{ij}$,
and $ K = K_{ij} h^{ij}$. Latin indices
$i,k,\ldots$ are 3-dimensional, they are raised and
lowered with the metric $ h_{ij}$ and its inverse $ h^{ij}$. For a general
introduction on this subject see, for example, the review article
\cite{Bartnik04b} and references therein.

If we think the initial data as a spacelike surface in the spacetime, with unit
timelike normal $t^\mu$, then the matter fields $\mu$ and $j^i$ are given in
terms of the energy momentum tensor $T_{\mu\nu}$ by
\begin{equation}
  \label{eq:17x}
  \mu=T_{\mu\nu} t^\mu t^\nu, \quad j_\nu =T_{\mu\nu}t^\nu.
\end{equation}
The dominant energy condition  \eqref{eq:15v}  implies
\begin{equation}
  \label{eq:65}
  \mu^2 \geq j_ij^i.
\end{equation}

The decomposition (\ref{eq:14}) of the matter fields translate to
\begin{equation}
  \label{eq:5x}
  \mu=\mu_{EM}+\mu_M, \quad j^i=j^i_{EM}+j^i_M,
\end{equation}
where we have defined 
\begin{equation}
  \label{eq:10v}
  \mu_{EM}=\frac{1}{4\pi}\left(E^iE_i+B^iB_i \right),\quad  j^i_{EM}=\epsilon^i{}_{jk} E^jB^k,
\end{equation}
where $\epsilon_{ilm}$ is the volume element of $h_{ij}$ and the electric field
$E$ and magnetic field $B$ are given by
\begin{equation}
  \label{eq:6x}
  E_\mu=F_{\mu\nu} t^\nu, \quad  B_\mu=- {}^*F_{\mu\nu} t^\nu,
\end{equation}
where ${}^*F_{\mu\nu}$ denotes the dual of $F_{\mu\nu}$  defined with respect to the volume element
$\epsilon_{\mu\nu\lambda\gamma}$ of the metric $g_{\mu\nu}$ by  the standard 
formula
\begin{equation}
  \label{eq:dual}
  {}^*F_{\mu\nu}=\frac{1}{2} F^{\alpha \beta}
    \epsilon_{\alpha \beta \mu  \nu}.
\end{equation}
The electric and magnetic fields satisfy  Maxwell constraint  equations
\begin{equation}
  \label{eq:29vc}
  D^i E_i=4\pi \rho, \quad   D^i B_i=0,
\end{equation}
where $\rho$ is the electric charge density. The relation between $\rho$ and
the spacetime electromagnetic current $\J^\mu$  is given  by $\rho=\J^\mu t_\mu$.

The initial data model an isolated system if the fields are weak far away from
sources. This physical idea is captured in the following definition of
asymptotically flat initial data set. In this article we assume that the
manifold $\Si$ is $\Rt$, hence the definition simplify slightly.  Consider
Cartesian coordinates $x^i$ with their associated euclidean radius
$r=\left( \sum_{i=1}^3 (x^i)^2 \right)^{1/2}$ and let $\delta_{ij}$ be the
euclidean metric components with respect to $x^i$.  The initial data set
$(\Si, h_{ij}, K_{ij}, \mu, j^i)$ is called \emph{asymptotically flat} if the
metric $h_{ij}$ and the tensor $K_{ij}$ satisfy the following fall off
conditions
\begin{equation}
  \label{eq:99}
  h_{ij}=\delta_{ij} +\gamma_{ij}, \quad K_{ij}=O(r^{-2}),
\end{equation}
where $\gamma_{ij}=O(r^{-1})$, $\partial_k \gamma_{ij}=O(r^{-2})$,
$\partial_l\partial_k\gamma_{ij}=O(r^{-3})$ and $\partial_k K_{ij}=O(r^{-3})$.
These conditions are written in terms of Cartesian coordinates $x^i$, here
$\partial_i$ denotes partial derivatives with respect to these coordinates.

We will assume that the initial data set has spherical symmetry. The
$\xi^i$ be one of the Killing vectors that generate the group $SO(3)$, then we
say the the initial data set is \emph{spherically symmetric} if
\begin{equation}
  \label{eq:31}
  \pounds_\xi h_{ij}=\pounds_\xi K_{ij}=\pounds_\xi \mu=\pounds_\xi j^i=0,
\end{equation}
for all the generators $\xi$ of $SO(3)$, where $\pounds$ denotes Lie
derivative. Note that we are imposing spherical symmetry also on the
sources. We also impose this condition on the electromagnetic field 
\begin{equation}
  \label{eq:32}
  \pounds_\xi E^i=\pounds_\xi B^i=\pounds_\xi \rho=\pounds_\xi  j^i_{EM} =0.
\end{equation}

There are several useful coordinates to describe spherically symmetric
metrics. In this article we will use the geodesic coordinates given by
\begin{equation}
\label{eq:metric}
 h = dl^2 + r^2(l)(d\theta ^2 + \sin^2\theta d \phi ^2),
\end{equation}
where $l$ is the proper radial distance to the center and $r(l)$ is the areal
radius. The function $r(l)$ is assumed to be smooth for $0\leq l
<\infty$. Regularity at the center  implies the
following conditions  for $r(l)$
\begin{equation}
  \label{eq:1}
  r(0)=0, \quad r'(0)=1,
\end{equation}
where the prime denotes derivative with respect to $l$. 
The asymptotically flat condition (\ref{eq:99}) implies
\begin{equation}
  \label{eq:2}
  \lim_{l\to \infty} r' = 1.
\end{equation}
The scalar curvature of the metric (\ref{eq:metric}) is given by
\begin{equation}
  \label{eq:21}
  R=-\frac{2}{r^2}\left(r'^2 +2rr'' -1  \right).
\end{equation}

Let $n^i$ denote the outwards unit normal vector to the spheres centered at the
origin, that is $ n=\partial/\partial l$.
The general form of the extrinsic curvature in spherical symmetric is given by   
\begin{equation}
\label{eq:extrinsic}
 K_{i j}= n_j n_j K_l + (g_{i j} - n_i n_j) K_r,
\end{equation}
where $K_l$ and $K_r$ are two functions of $l$. The asymptotically flat condition
(\ref{eq:99}) implies
\begin{equation}
  \label{eq:33}
   \lim_{l\to \infty} rK_r=0.
\end{equation}

Using (\ref{eq:21}) and (\ref{eq:extrinsic}) we can write the constraint
equations (\ref{const1})--(\ref{const2}) in spherically symmetric in the
following form
\begin{align}
  K_r \left( K_r +2 K_l \right) - \frac{1}{r^2} \left( {r'}^2 + 2 r r'' -1
  \right) &= 8 \pi \mu,  \label{eq:hamiltonean}   \\
   {K_r}' + \frac{r'}{r} \left( K_r - K_l \right) &= 4 \pi j, \label{eq:momentum}
\end{align}
where $j$ is the radial component of the current density $j=j^in_i$, which is
the only non-trivial component due to the spherical symmetry.  The dominant
energy condition is given by
\begin{equation}
  \label{eq:3}
  \mu \geq |j|. 
\end{equation}
Let $E=E^in_i$ and $B=B^in_i$, then equations (\ref{eq:29vc}) are given by 
\begin{equation}
 \label{maxwell}
   \frac{1}{r^2} (  E r^2)' = 4\pi \rho, \quad \frac{1}{r^2} (  B r^2)'  = 0,
\end{equation}
where $\rho$ is the electric charge density. The energy density $\mu$ is given by
\begin{equation}
  \label{eq:4v}
  \mu=\mu_M + \frac{1}{8\pi}\left(E^2+B^2\right).
\end{equation}
Note that since $B^i$ are $E^i$ are radial then $j^i_{EM}=0$ and hence the
current density $j^i$ has no electromagnetic contribution in spherical
symmetry.  We say the data is \emph{electrovacuum} if $\mu_M=0$ and $j=0$.

The electric charge contained in $\B$ is given by
\begin{equation}
  \label{eq:5}
  Q=4\pi \int_0^{l_0} \rho  r^2  \,dl. 
\end{equation}
Using Gauss theorem and equation (\ref{maxwell}) we obtain that the charge can also be written as
\begin{equation}
  \label{eq:6}
  Q= E r^2. 
\end{equation}

Finally,  the outgoing future and past null expansions are given by
\begin{equation}
\label{eq:expansions}
\theta ^{+} = \frac{2}{r} \left( r' + K_r r \right), \quad  \theta ^{-} =
\frac{2}{r} \left( r' - K_r r \right).
\end{equation}


\end{document}